\documentclass[aps,preprint]{revtex4}%
\usepackage{amsfonts}
\usepackage{amsmath}
\usepackage{amssymb}
\usepackage{graphicx}%
\begin{document}
\preprint{ }
\title[ ]{\textbf{Thermo-magnetic Effects in an External Magnetic Field in the
Logarithmic-Quark Sigma Model}}
\author{\textbf{M. Abu-Shady}}
\affiliation{{\small Department of Applied Mathematics, Faculty of Science, Menoufia
University, Egypt}}
\author{}
\affiliation{}
\author{}
\affiliation{}
\affiliation{}
\keywords{Mean-field approximation, Quark sigma model, Magnetic Catalysis}
\pacs{PACS number}

\begin{abstract}
The phenomenon of magnetic catalysis of chiral symmetry breaking in the
quantum chromodynamic theory in the framework of logarithmic quark sigma model
is studied. Thermodynamic properties are calculated in the mean-field
approximation such as the pressure, the entropy density, the energy density,
and measure interaction. The pressure, the entropy density, and the energy
density increase with increasing temperature and an external magnetic field.
The critical temperature increases with increasing an external magnetic field.
In addition, the chiral phase transition is crossover in the presence of an
external magnetic field with absent of baryonic chemical potential when
explicit symmetry breaking is included. A comparison is presented with the
original sigma model and other works. A conclusion indicates that the
logarithmic quark model enhances the magnetic catalysis phenomenon.

\textbf{Keywords:} Chiral Lagrangian density, Magnetic catalysis, Mean-field approximation

\end{abstract}
\volumeyear{ }
\volumenumber{ }
\issuenumber{ }
\eid{ }
\startpage{1}
\endpage{ }
\maketitle

\section{Introduction}

The study of the influence of an external magnetic fields on the fundamental
properties of quantum chromodynamic (QCD) theory such as the confinement of
quark and gluons at low energy and asymptotic freedom at high energy is still
a matter of great theoretical and experimental $\left[  1\right]  $. The
transition from composite objects to colored quarks and gluons has several
characteristics of both theoretical and phenomenological relevance. One such
characteristic is the equation of state which is the fundamental relation
encoding the thermodynamic properties of the system $\left[  1-4\right]  $. So
for, most estimates have been carried out at vanishing chemical potential with
the aid of effective theories such as the linear sigma model (LSM) $\left[
5\right]  $ and the Numbu-Jona-Lasinio (NJL) model $\left[  6,7\right]  $ in
the mean field approximation.

The linear sigma model exhibits many of global symmetries of QCD theory. The
model was originally introduced by Gell-Mann and Levy $\left[  8\right]  $
with the purpose of describing pion-nucleon interactions. During the last
years an impressive amount of work has been done with this model. The idea is
to consider it as an effective low energy approach for QCD theory that the
model has some aspects of QCD theory such as the chiral symmetry. Thus, the
model is successfully to describe the most of hadron properties at low energy
such as in Refs. $\left[  5,9,10\right]  $. Some observables that are
calculated in this model are conflict with experimental data. So researchers
interest to modify this model to provide a good description of hadron
properties such as in Refs. $\left[  11-15\right]  .$

In Ref. $\left[  16\right]  ,$ the authors have modified the linear sigma
model by including the logarithmic mesonic potential and study its effect on
the phase transition at finite temperature. In addition, the comparison with
other models is done. On the same hand, the logarithmic sigma model
successfully describes nucleon properties at finite temperature and chemical
potential $\left[  17\right]  $.

To continue the investigation that started in Ref. $\left[  18\right]  $. In
this paper, we investigate the effect of external magnetic field on the
thermodynamic properties in the framework of logarithmic quark sigma model at
finite temperature and chemical potential. So for no attempts have done to
investigate the thermodynamic properties in the framework of the logarithmic
sigma model.

This paper is organized as follows: The original sigma model is briefly
presented in Sec. 2. Next, the effective logarithmic mesonic potential in the
presence of external magnetic field is presented in Sec. 3. The results are
discussed and are compared with other works in Secs. 4 and 5, respectively.
Finally, the summary and conclusion are presented in Sec. 6.

\section{The chiral sigma model with original effective potential}

The interactions of quarks via the exchange of $\sigma-$ and $\mathbf{\pi}$ -
meson fields are given by the Lagrangian density $[5]$ as follows:
\begin{equation}
{\small L}\left(  r\right)  {\small =i}\overline{\Psi}{\small \gamma}^{\mu
}{\small \partial}_{\mu}{\small \Psi+}\frac{1}{2}\left(  \partial_{\mu}%
\sigma\partial^{\mu}\sigma+\partial_{\mu}\mathbf{\pi}.\partial^{\mu
}\mathbf{\pi}\right)  {\small +g}\overline{\Psi}\left(  \sigma+i\gamma
_{5}\mathbf{\tau}.\mathbf{\pi}\right)  {\small \Psi-U}_{1}{\small (\sigma
,\pi),} \tag{1}%
\end{equation}
with
\begin{equation}
{\small U}_{1}{\small (\sigma,\pi)=}\frac{\lambda^{2}}{4}\left(  \sigma
^{2}+\mathbf{\pi}^{2}-\nu^{2}\right)  ^{2}+{\small m}_{\pi}^{2}{\small f}%
_{\pi}{\small \sigma.} \tag{2}%
\end{equation}
where $U\left(  \sigma,\mathbf{\pi}\right)  $ is the meson-meson interaction
potential and $\Psi,\sigma$ and $\mathbf{\pi}$ are the quark, sigma, and pion
fields, respectively. In the mean-field approximation, the meson fields are
treated as time-independent classical fields. This means that we replace the
power and products of the meson fields by corresponding powers and the
products of their expectation values. The meson-meson interactions in Eq. (2)
lead to hidden chiral $SU(2)\times SU(2)$ symmetry with $\sigma\left(
r\right)  $ taking on a vacuum expectation value \ \ \ \ \ \ \
\begin{equation}
\ \ \ \ {\small \ \ }\left\langle \sigma\right\rangle {\small =-f}_{\pi
}{\small ,} \tag{3}%
\end{equation}
where $f_{\pi}=93$ MeV is the pion decay constant. The final \ term in Eq. (2)
included to break the chiral symmetry explicitly. It leads to the partial
conservation of axial-vector isospin current (PCAC). The parameters
$\lambda^{2}$and $\nu^{2}$ can be expressed in terms of$\ f_{\pi}$, sigma and
pion masses as,
\begin{equation}
{\small \lambda}^{2}{\small =}\frac{m_{\sigma}^{2}-m_{\pi}^{2}}{2f_{\pi}^{2}%
}{\small ,} \tag{4}%
\end{equation}%
\begin{equation}
{\small \nu}^{2}{\small =f}_{\pi}^{2}{\small -}\frac{m_{\pi}^{2}}{\lambda^{2}%
}{\small .} \tag{5}%
\end{equation}

\section{The effective logarithmic potential in the presence of magnetic
field}

In this section, the logarithmic mesonic potential $U_{2}(\sigma,\pi)$ is
applied. In Eq. (6), the logarithmic potential is included with the external
magnetic field at finite temperature and baryonic chemical potential $\left[
2\right]  $ as follows,%

\begin{equation}
U_{eff}(\sigma,\pi)=U_{2}(\sigma,\pi)+U_{Vaccum}+U_{Matter}+U_{Medium},
\tag{6}%
\end{equation}
where%
\begin{equation}
\ \ \ \ \ \ U_{2}\left(  \sigma,\mathbf{\pi}\right)  =-\lambda_{1}^{2}\left(
\sigma^{2}+\mathbf{\pi}^{2}\right)  +\lambda_{2}^{2}\left(  \sigma
^{2}+\mathbf{\pi}^{2}\right)  ^{2}\log\left(  \frac{\sigma^{2}+\mathbf{\pi
}^{2}}{f_{\pi}^{2}}\right)  +m_{\pi}^{2}f_{\pi}\sigma, \tag{7}%
\end{equation}
In Eq. 7, the logarithmic potential satisfies the chiral symmetry when
$m_{\pi}\longrightarrow0$ as well as in the original potential in Eq. 2.
Spontaneous chiral symmetry breaking gives a nonzero vacuum expectation for
$\sigma$ and the explicit chiral symmetry breaking term in Eq. 7 gives the
pion its mass.%
\begin{equation}
\left\langle \sigma\right\rangle =-f_{\pi}. \tag{8}%
\end{equation}
Where%
\begin{equation}
\ \ \ \ \ \lambda_{1}^{2}=\frac{m_{\sigma}^{2}-7m_{\pi}^{2}}{12}, \tag{9}%
\end{equation}%
\begin{equation}
\ \ \ \ \ \lambda_{2}^{2}=\frac{m_{\sigma}^{2}-m_{\pi}^{2}}{12f_{\pi}^{2}}.
\tag{10}%
\end{equation}
For details, see Refs. $\left[  17,18\right]  $. To include the external
magnetic field in the present model, we follow Ref. $\left[  2\right]  $ by
including the pure fermionic vacuum contribution in the free potential energy.
Since this model is renormalizable the usual procedure is to regularize
divergent integrals using dimensional regularization and to subtract the ultra
violet divergences. This procedure gives the following result%
\begin{equation}
{\small U}_{Vac%
\operatorname{u}%
m}=\frac{N_{c}N_{f}~g^{4}}{(2\pi)^{2}}(\sigma^{2}+\pi^{2})^{2}(\frac{3}{2}%
-\ln(\frac{g^{2}(\sigma^{2}+\pi^{2})}{\Lambda^{2}})), \tag{11}%
\end{equation}
where $N_{c}=3$ and $N_{f}=2$ are color and flavor degrees of freedom,
respectively and $\Lambda$ is mass scale,
\begin{equation}
U_{Matter}=\frac{N_{c}}{2\pi^{2}}\sum_{f~=u}^{d}(\left\vert q_{f}\right\vert
B)^{2}[\zeta^{(1,0)}(-1,x_{f})-\frac{1}{2}(x_{f}^{2}-x_{f})\ln x_{f}%
+\frac{x_{f}^{2}}{4}] \tag{12}%
\end{equation}
In Eq. 12, we have used $x_{f}=\frac{g^{2}(\sigma^{2}+\pi^{2})}{(2\left\vert
q_{f}\right\vert B)}$ and $\zeta^{(1,0)}(-1,x_{f})=\frac{d~\zeta(z,x_{f})}%
{dz}\mid_{z~=-1}$ that represents the Riemann-Hurwitz function, and also
$\left\vert q_{f}\right\vert $ is the absolute value of quark electric charge
in the external magnetic field with intense $B$.%

\begin{equation}
U_{Medium}=\frac{N_{c}}{\left(  2\pi\right)  ^{2}}\sum_{f~=u}^{d}\sum
_{k=0}^{\infty}\alpha_{k}(\left\vert q_{f}\right\vert B)%
{\displaystyle\int\limits_{\infty}^{-\infty}}
dp_{z}[T\ln\left[  1+e^{-\frac{^{[E_{P,k}(B)+\mu]}}{T}}\right]  +T\ln\left[
1+e^{-\frac{^{[E_{P,k}(B)-\mu]}}{T}}\right]  ]. \tag{13}%
\end{equation}
where $E_{P,k}(B)=\sqrt{P_{z}^{2}+2k\left\vert q_{f}\right\vert B+M^{2}}$, $M$
is the effective self-consistent quark mass and $\mu$ is baryonic chemical potential.

\section{ Discussion of results}

In this section, we study the effective potential of the logarithmic sigma
model. For this purpose, we numerically calculate the effective potential in
Eq. (6). The parameters of the present model are the coupling constant $g$ and
the sigma mass $m_{\sigma}$. The choice of free parameters of coupling
constant $g$ and sigma mass $m_{\sigma}$ based on Ref. $\left[  2\right]  .$
The parameters are usually chosen so that the chiral symmetry is spontaneously
broken in the vacuum and the expectation values of the meson fields. In this
work, we consider set of parameters, $m_{\pi}=138$ MeV, $m_{\sigma}=600$ MeV,
and $g=4.5$. We start by calculating the quantities such as the dimensionless
of pressure $\frac{P_{n}}{T^{4}}$ , the dimensionless energy density
$\frac{E_{n}}{T^{3}}$, and the dimensionless entropy density $\frac{S}{T^{3}}%
$. \ These quantities can be readily obtained by the effective potential that
defines in Eq. $\left(  6\right)  $, gives the negative pressure
$U_{eff}(\sigma,\pi)=-P_{n}$, then the net quark number density is obtained
from $\rho=\frac{dP_{n}}{d\mu}$, and the entropy density from $s=\frac{dP_{n}%
}{dT}$ while the energy density is $E_{n}=-P_{n}+TS+\mu\rho$ see Ref. $\left[
2\right]  $. These quantities are displayed in Figs. (1, 2, 3, 4, 5, and 6).
First of all, one observes that the quantities are calculated at zero and
finite strong magnetic field in the two cases at vanishing of chemical
potential and non-vanishing of chemical potential.%

\begin{center}
\includegraphics[
height=5.047in,
width=5.047in
]%
{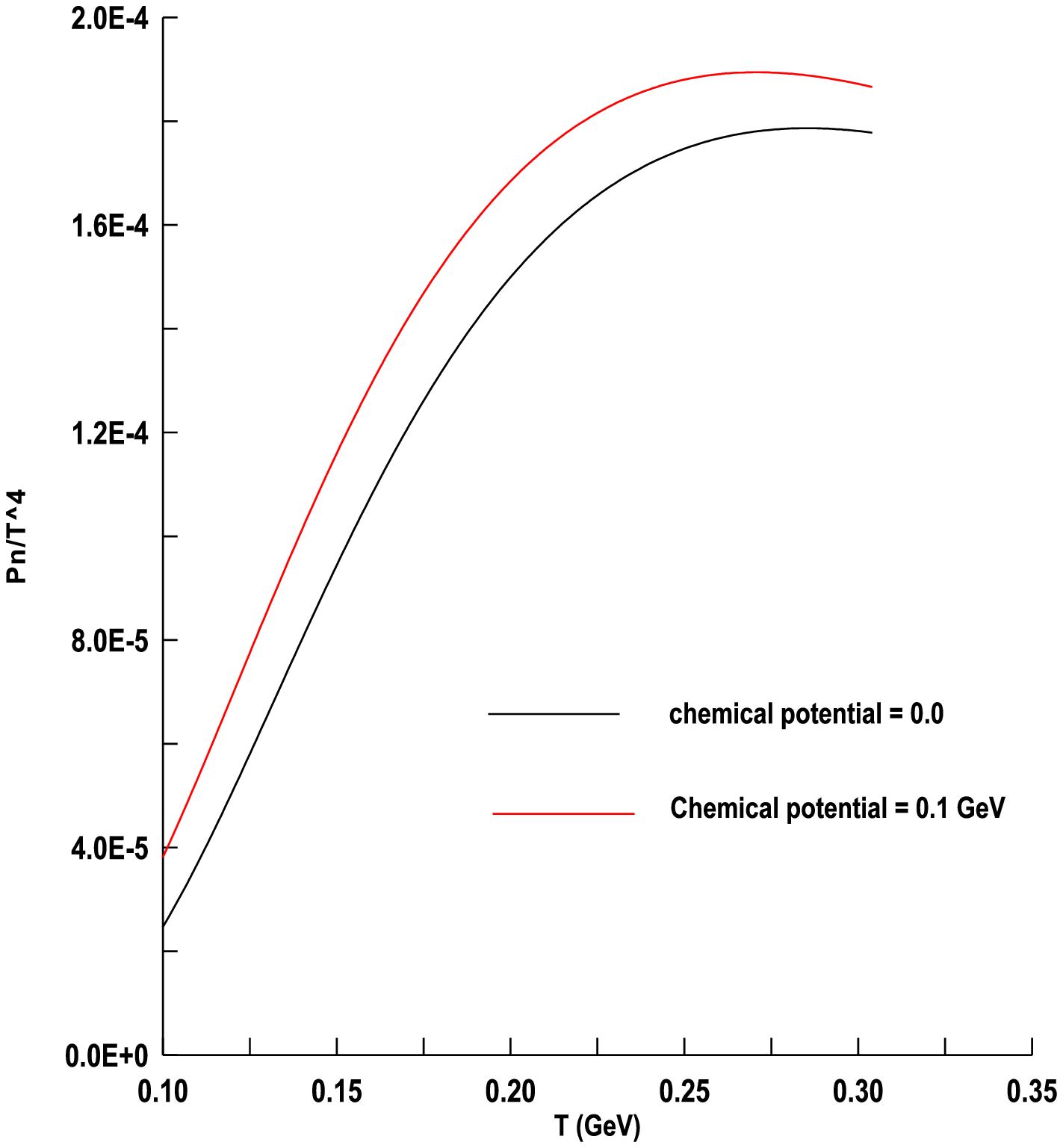}%
\\
{\small Fig. 1}\textbf{. }{\small The dimensionless pressure is plotted as a
function of temperature at }$\mu=0~${\small  \ and }$\mu=0.1$ {\small GeV at
vanishing magnetic field }%
\end{center}
%

\begin{center}
\includegraphics[
height=5.047in,
width=5.047in
]%
{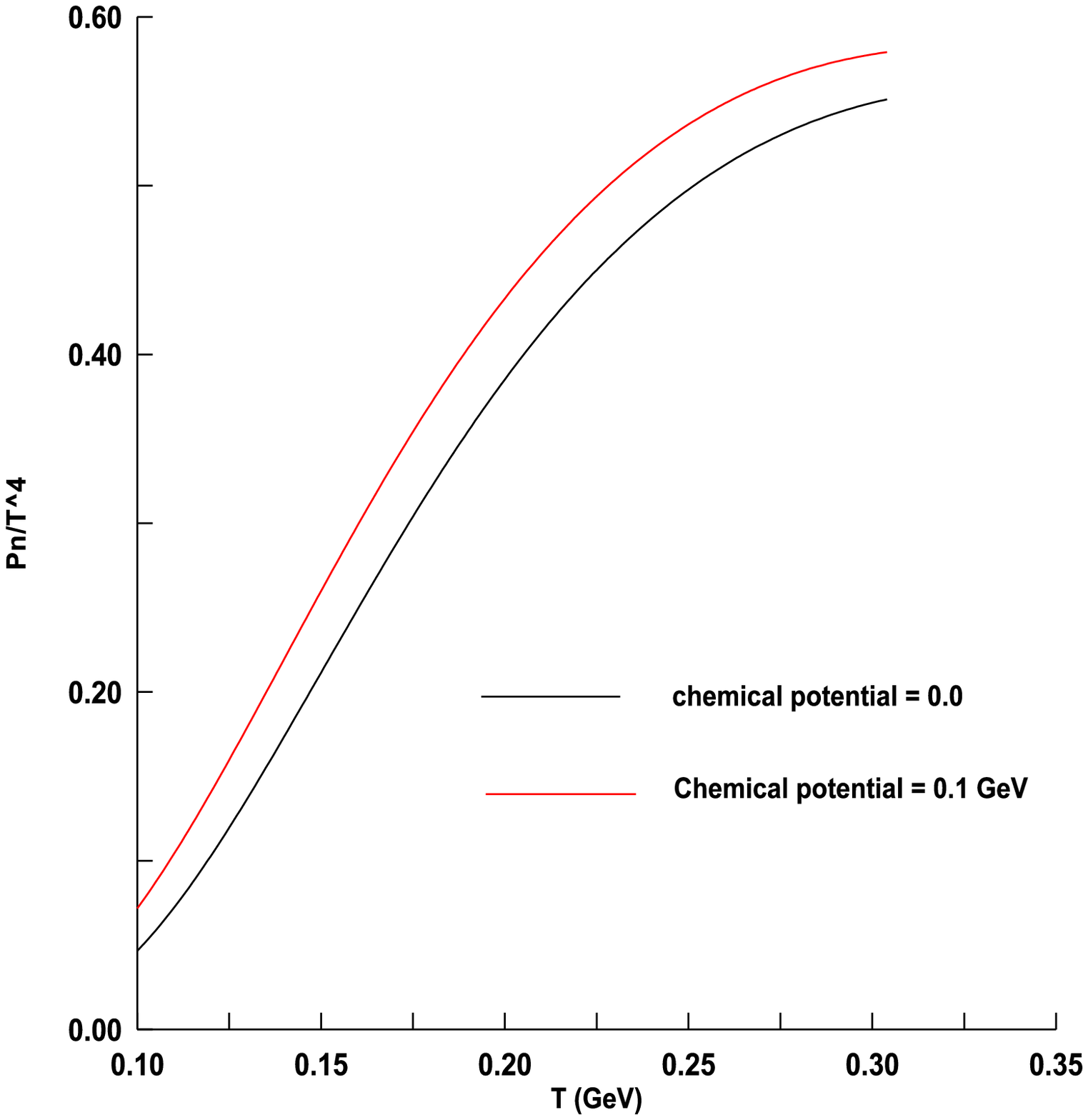}%
\\
{\small Fig. 2. The dimensionless pressure is plotted as a function\ of
temperature for at }$\mu=0${\small  \ and }$\mu=0.1${\small  GeV at strong
magnetic field eB = 0.4 GeV}$^{2}.$%
\end{center}
\ \ \ \ \ \ \ \ \ \ \ 

In Fig. 1, the dimensionless pressure is plotted as a function of temperature
at vanishing of magnetic field $(B)$. At vanishing of chemical potential
($\mu=0)$, the pressure increases with increasing temperature. The behavior is
good qualitative agreement with recent lattice calculation in Ref. $\left[
1\right]  $, in which the generalized integral method is used and the pressure
increases with increasing temperature. Also, the behavior of pressure is in
qualitative agreement the original sigma model and NJL model $\left[
2\right]  $. On the same hand, the behavior of pressure is qualitative
agreement work of Tawfik et al. $\left[  3\right]  ,$ in which the hadron gas
and the Polyakov linear sigma models are applied in their calculations. At
$\mu\neq0$, we note that the curve shifts to higher values, in particular, at
higher-values of temperature. At finite baryonic chemical potential is not
considered in many works such as in Ref. $\left[  3\right]  $. In Fig. 2, the
effect of strong magnetic field is studied on the behavior of pressure, then
we choose magnetic field\ $eB=0.2$ GeV$^{2}$ as in Refs. $\left[
1,2,3\right]  $. We note that qualitative agreement between Fig. 1 represents
the case of vanishing magnetic field and Fig. 2 represents the presence of an
external magnetic field. The effect of magnetic field appears on the pressure
value that the pressure increases strongly at higher-values of temperature.
\ in comparison with vanishing magnetic field as in Fig. 1. This behavior is
qualitative agreement with Refs. $\left[  1,3\right]  $.%

\begin{center}
\includegraphics[
height=5.047in,
width=5.047in
]%
{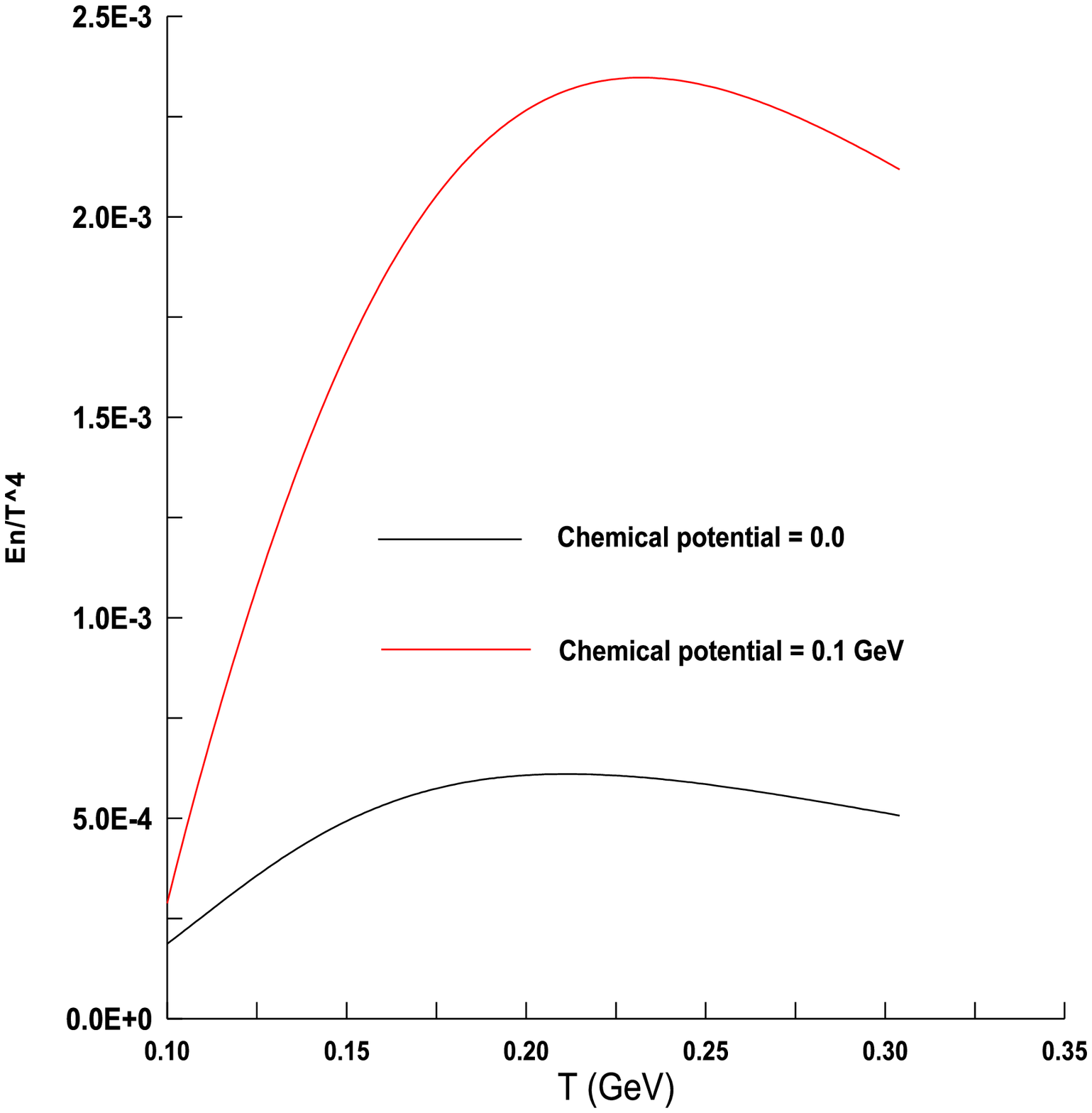}%
\\
{\small Fig. 3}\textbf{. }{\small The dimensionless energy density is plotted
as a function of temperature for at }$\mu=0${\small  \ and }$\mu=0.2${\small
GeV at vanishing magnetic field }%
\end{center}
%

\begin{center}
\includegraphics[
height=5.047in,
width=5.047in
]%
{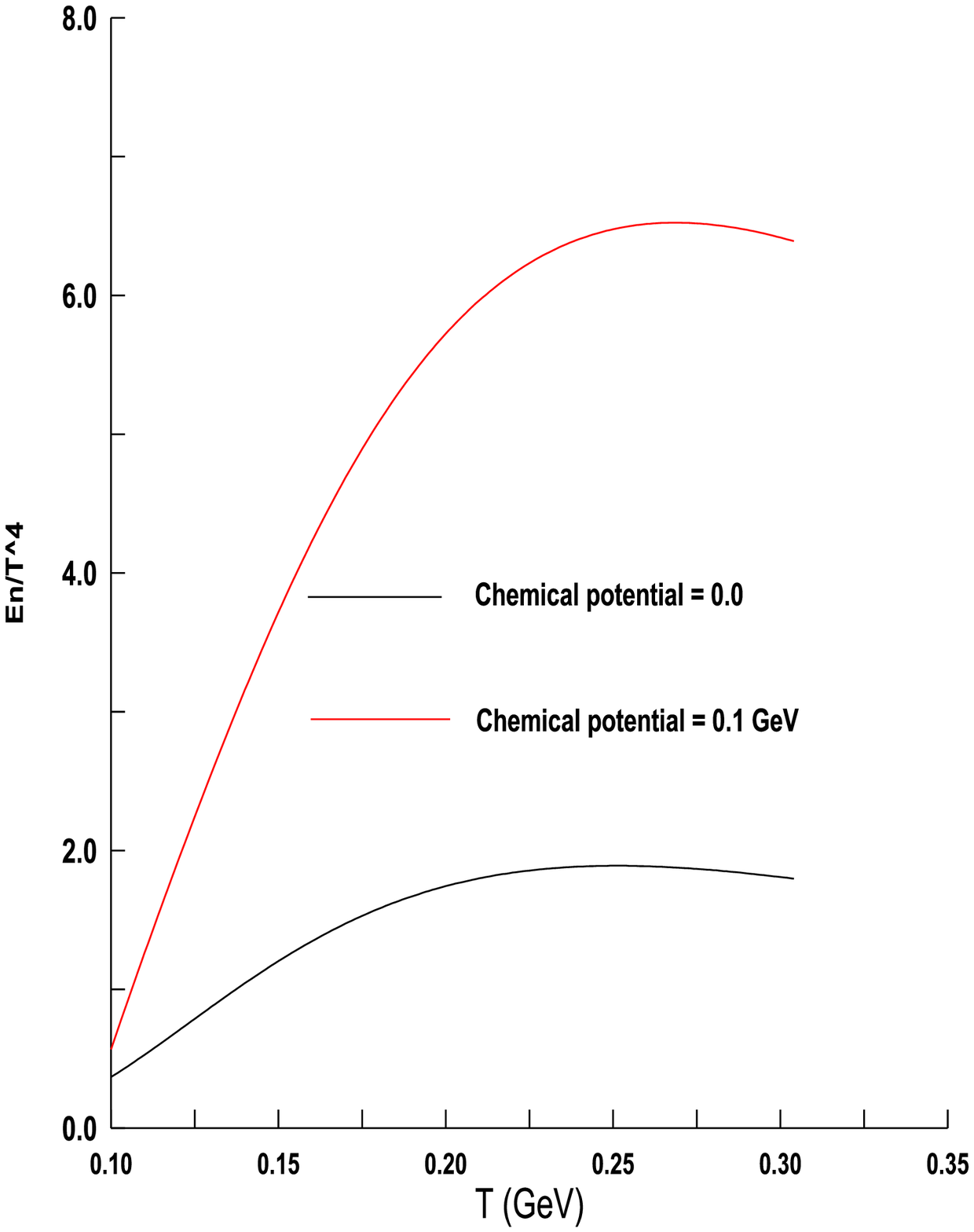}%
\\
{\small Fig. 4}\textbf{. }{\small The dimensionless energy density is plotted
as a function of \ temperature at }$\mu=0~${\small  \ \ ~and }$\mu
=0.1${\small ~GeV at non-vanishing magnetic field eB =0.4 GeV}$^{2}$%
\end{center}

In Fig. 3, the dimensionless energy density is plotted as a function of
temperature. We note that the energy density increases with increasing
temperature at vanishing and nonvanishing chemical potential. At vanishing of
baryonic chemical potential ($\mu=0)$, the energy density increases with
increasing temperature up to 0.225 GeV and then slowly decreases at larger
values of temperature above critical temperature due to taking ratio
$\frac{E_{n}}{T^{4}}$. By increasing magnetic field as in Fig 4, the behavior
of the energy density is a qualitative agreement with the behavior at
vanishing magnetic field as in Fig. 3 and supporting with Refs. $\left[
1,3\right]  .$%

\begin{center}
\includegraphics[
height=5.047in,
width=5.047in
]%
{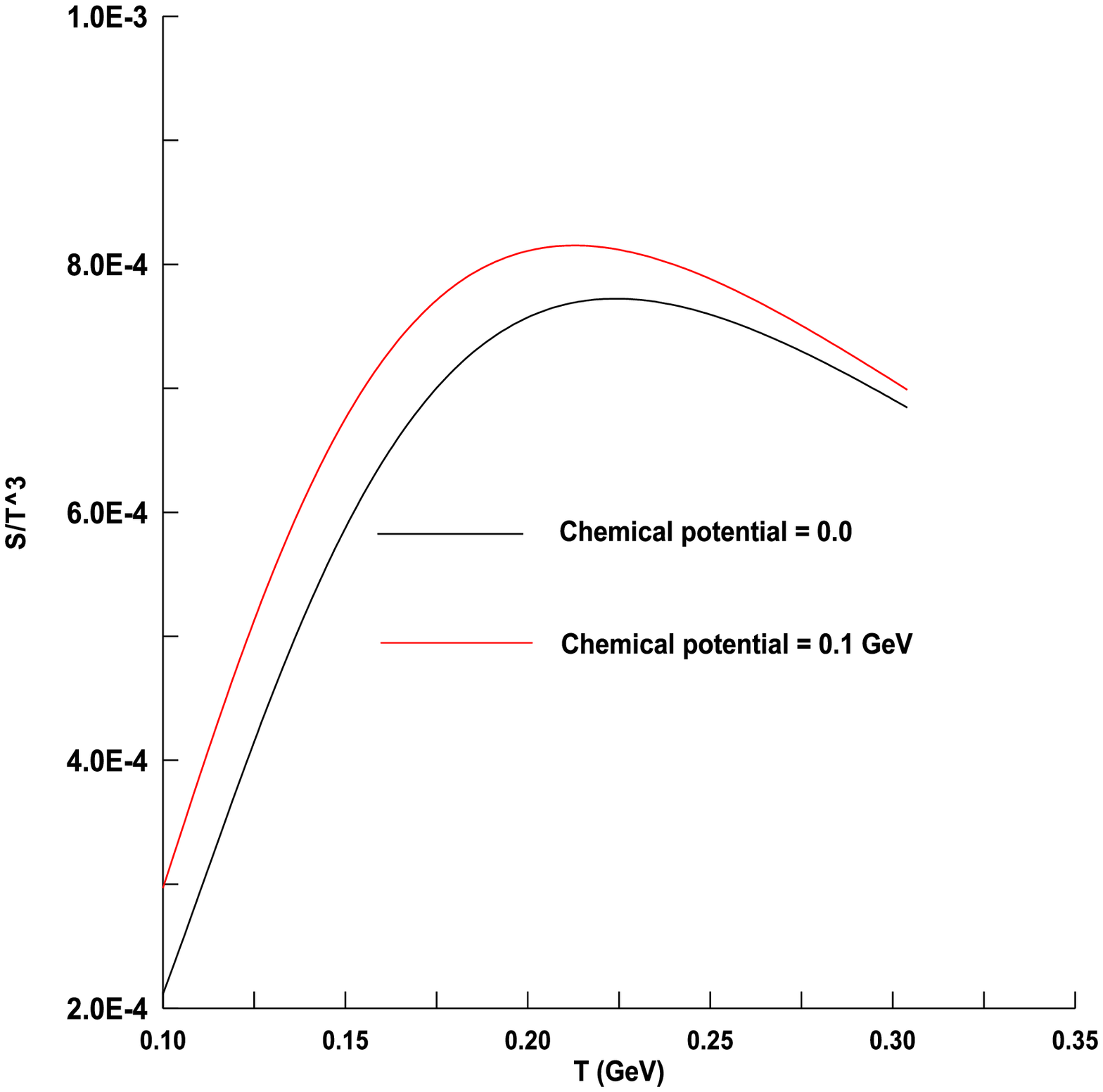}%
\\
{\small Fig. 5}\textbf{. }{\small The dimensionless entropy density is plotted
as a function of \ temperature for at }$\mu=0${\small  \ and }$\mu
=0.1${\small  GeV at vanishing magnetic field }%
\end{center}
%

\begin{center}
\includegraphics[
height=5.047in,
width=5.047in
]%
{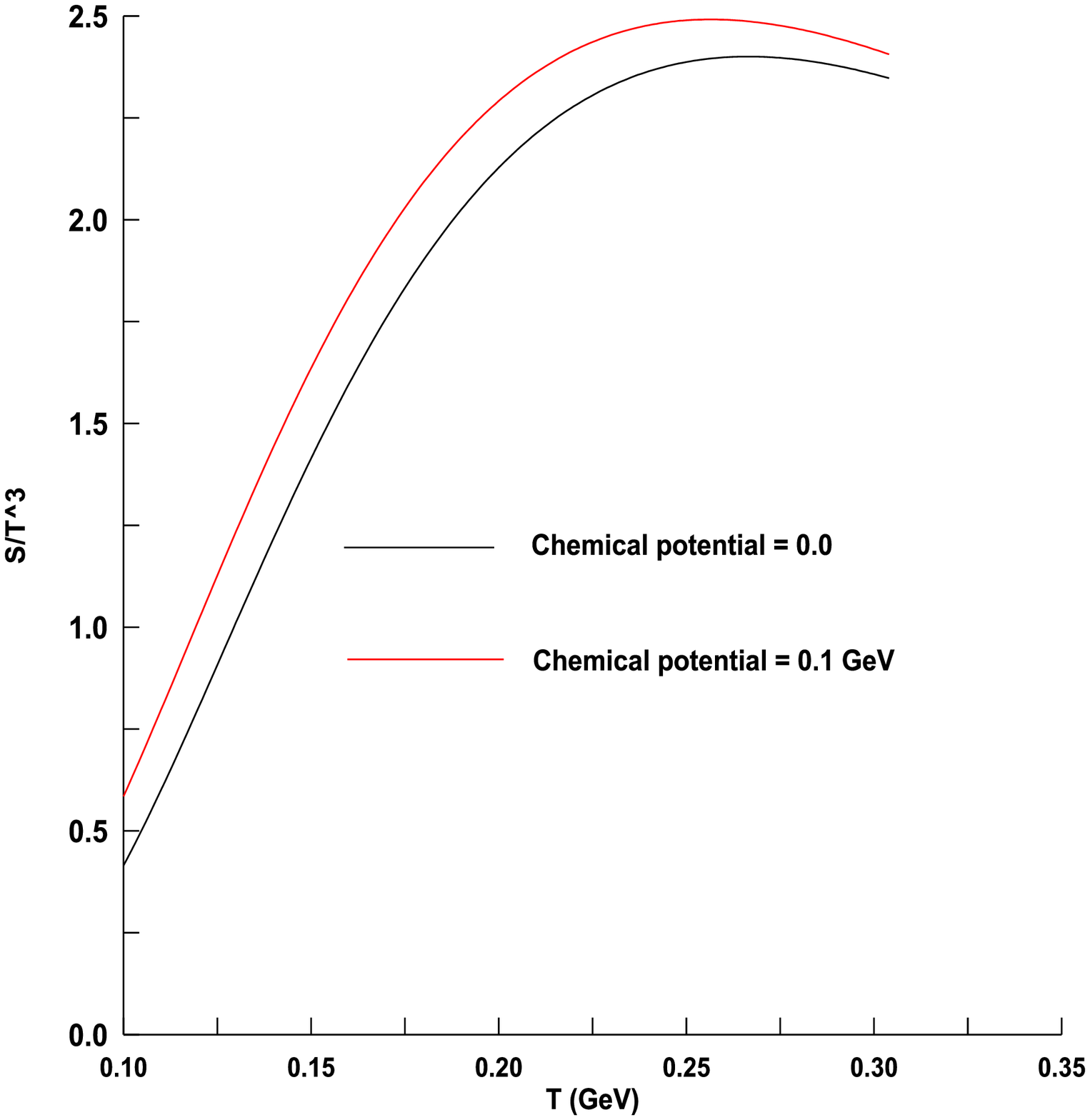}%
\\
{\small Fig. 6}\textbf{. }{\small The dimensionless entropy density is plotted
as a function of \ temperature for at }$\mu=0${\small  \ and }$\mu
=0.1${\small  \ GeV at magnetic field eB = 0.4 GeV}$^{2}$%
\end{center}

In Fig. 5, the dimensionless entropy density is plotted as a function of
temperature. The entropy density increases with increasing temperature at
vanishing and nonvanishing chemical potential. The effect of temperature
appears on entropy density at higher values of temperatures. By increasing
magnetic field as in Fig. 6, the behavior of the entropy density is a
qualitative agreement with the behavior at vanishing magnetic field as in Fig.
5. Also, we note that entropy density decreases at higher temperature above
critical temperature $T_{c}=0.2$ GeV. This due to taking ratio $\frac{S}%
{T^{3}}$%

\begin{center}
\includegraphics[
height=5.047in,
width=5.047in
]%
{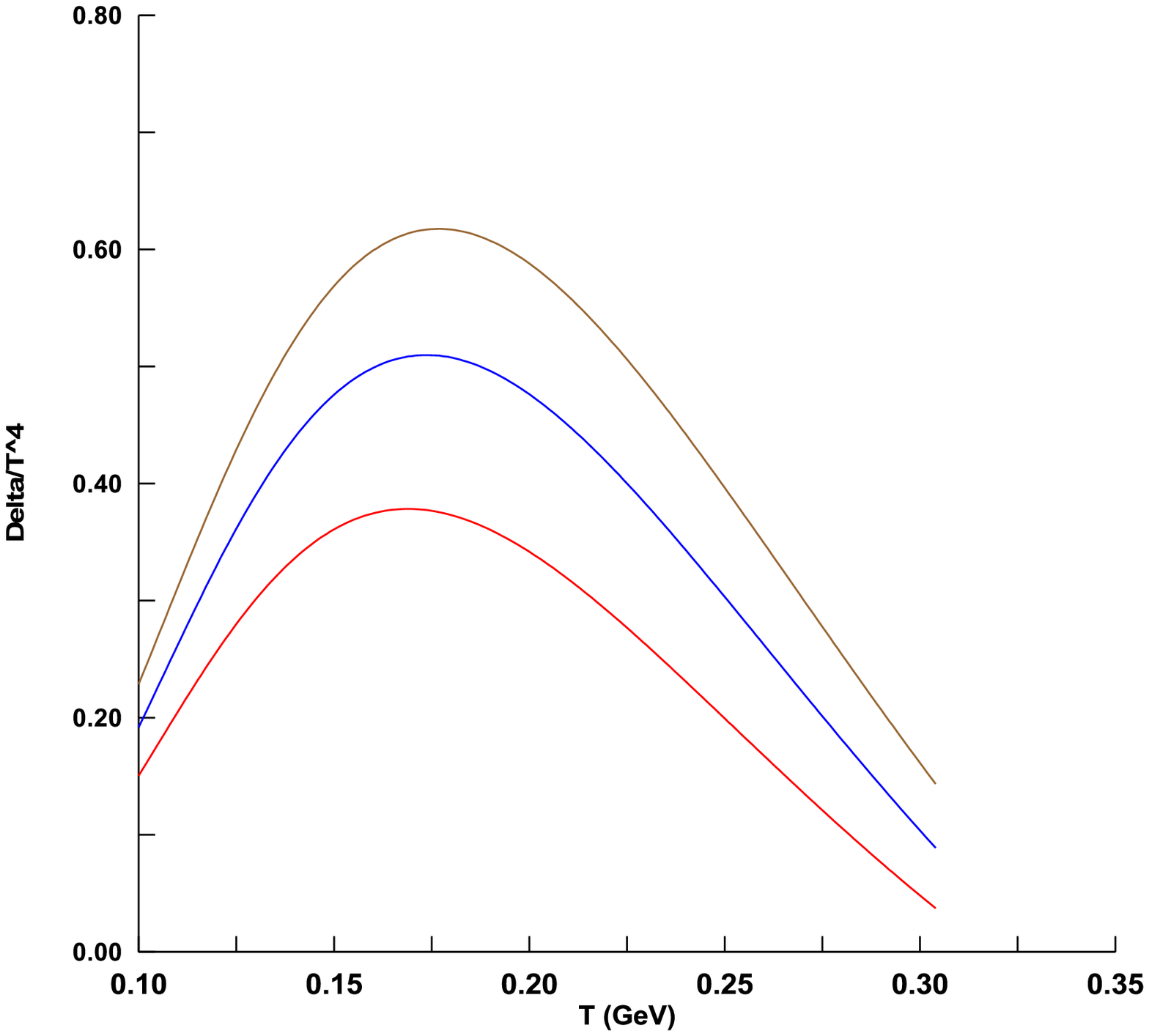}%
\\
{\small Fig. 7. The dimensionless interaction measure is plotted as a function
of temperature for \ different values of magnetic field eB = 0.2, 0.3, \ and
0.4 GeV}$^{2}${\small  are arranged from down to up. }%
\end{center}
%

\begin{center}
\includegraphics[
height=5.047in,
width=5.047in
]%
{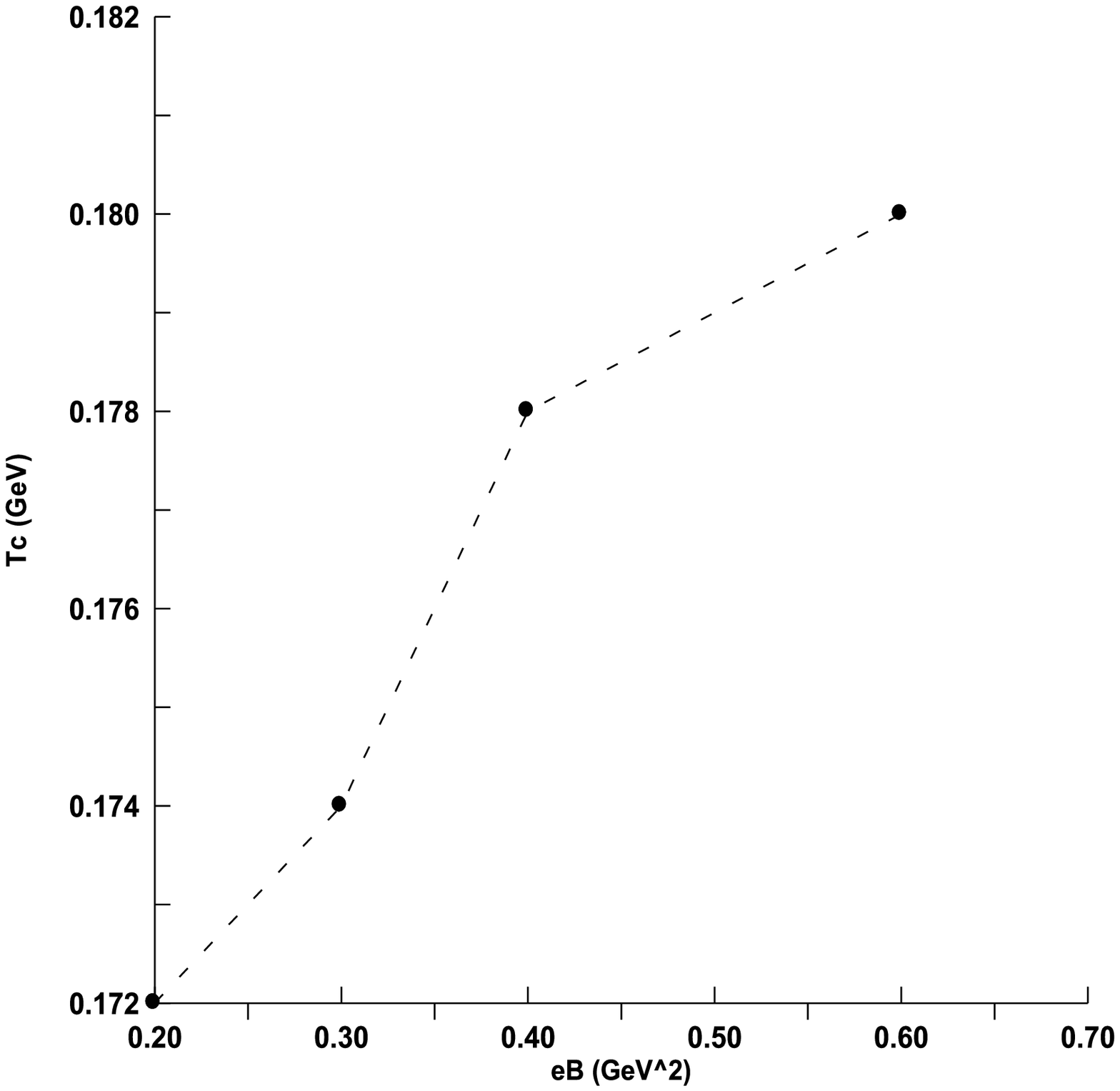}%
\\
{\small Fig. 8}\textbf{. }{\small Critical of temperature is plotted as a
function of magnetic field at vanishing of \ baryonic chemical potential.}%
\end{center}

In Fig. $\left(  7\right)  $, the measure interaction is plotted as a function
of temperature. The measure interaction increases with increasing temperature
up to the top curve and then gradually decreases with increasing temperature.
By increasing magnetic field, we note that the top shifts in direction of
increasing temperature. This behavior is an agreement with original sigma
model and NJL model $\left[  2\right]  $. This behavior is interpreted as
crossover phase transition. In Fig. (8), the critical temperature increases
with increasing magnetic field. Thus, the logarithmic quark enhances magnetic
catalysis as well as in the original sigma model and NJL model $\left[
2\right]  $. By including Polyakov loop to take confinement into account. The
results show that $T_{c}$ related to deconfinment also increases with
increasing magnetic field $\left[  26\right]  $. On other hand, the first
lattice attempt to solve the problem taken two quark flavors and high values
of pion mass $(m_{\pi}=200-400$ MeV) confirming that $T_{c}$ increasing with
increasing magnetic field $\left[  27\right]  $.

\section{Summary and conclusion}

In this work, we have employed the effective logarithmic potential to study
thermodynamic properties in the presence of an external magnetic field. So,
the novelty in this work, thermodynamic properties are investigated in the
framework of logarithmic quark sigma model. The present results are agreement
with first lattice calculations as in Ref. $\left[  27\right]  $ and original
sigma model and NJL model as in Refs. $\left[  2\right]  $. In addition, the
present calculations are carried out beyond the zero chemical potential which
are not taken many recent works. Our conclusion indicates that the logarithmic
quark sigma model enhances magnetic catalysis at finite temperature and
baryonic chemical potential.

\section{\textbf{References}}

\begin{enumerate}
\item G. S. Bali, F. Bruckmann, G. Endr\H{o}di, S. D.Katz, A. Schafer, J. High
Energy Phys. \textbf{177}, 35 (2014)

\item G. N. Ferrari, A. F. Garcia, and M. B. Pinto, Phys. Rev. D \textbf{86},
096005 (2014).

\item A. N. Tawfik, A. M. Diab, N. Ezzelarab, A. G. Shalaby, Advances in High
Energy Physics \textbf{2016}, 1381479 (2016).

\item R. L. S. Farias,V. S. Timoteo, S. S. Avancini, M. B. Pinto, G. Krein,
hep-ph%
$\backslash$%
1603.03847 (2016).

\item M. Birse and M. Banerjee, Phys. Rev. D \textbf{31}, 118 (1985).

\item R. Gatto and M. Ruggier D \textbf{83}, 040163 (2011).

\item S. S. Avancini, D. P. Menezes and C. Providencia, C \textbf{83}, 065805 (2011).

\item M. Gell-Mann and M. Levy, Nuono Cinmento \textbf{16}, 705 (1960).

\item M. Abu-Shady, Inter. J. Mod. Phys. A \textbf{26}, 235 (2011)

\item T.S.T. Aly, M. Rashdan, and M. Abu-Shady, Inter. J. Theor. Phys.
\textbf{45}, 1645 (2006).

\item M. Abu-Shady and M. Soleiman, Phys. Part. and Nuclei Lett. \textbf{10},
683 (2013)

\item M. Abu-Shady, Inter. J. Theor. Phys. \textbf{48} (4), (2009)

\item M Abu-Shady, Inter. J. of Mod. Phys. E \textbf{21} (06), 1250061 (2012)

\item M. Abu-Shady, Inter. J. Mod. Phys. A \textbf{26}, 235 (2011).

\item M. Abu-Shady, Mod. Phys. Lett. A \textbf{29}, 1450176 (2014).

\item M Abu-Shady, Inter. J. Theor. Phys. 48, 115-126 (2009)

\item M. Abu-Shady and A. Abu-Nab, the Euro. Phys. J. Plus \textbf{130}, 248 (2015).

\item M. Abu-Shady, Applied Math. and Information Sciences Lett. \textbf{4},
\ 5 (2016)

\item A. Goyal and M. Dahiya, Phys. Rev. D \textbf{62}, 025022 (2011).

\item S. P. Klevansky and R. H. Lemmar, Phys. Rev. D \textbf{39}, 3478 (1989).

\item I. A. Shushpanov and A. V. Smilga, Phys. Lett. B \textbf{16}, 402 (1997).

\item I. A. Shushpanov and A. V. Smilga, Phys. Lett. B \textbf{16}, 351 (1997).

\item H. Suganuma and T. Tastsumi, Annals. Phys. \textbf{208}, 470 (1991).

\item K. G. Klimenko and T. Mat. Fiz. \textbf{89}, 211 (1991).

\item V. P. Gusynin, V. A. Miransky, and I. A. Shovkovy, Phy. Rev. Lett.
\textbf{73}, 3499 (1994).

\item A. J. Mizher, M. N. Chernoub, and E. S. Fraga, Phys. Rev. D \textbf{82},
105016 (2010).

\item M. D. Elia, S. Mukherjee and F. Sanfilippo, Phys. Rev. D \textbf{82},
051501 (2010).
\end{enumerate}

\end{document}